\documentclass[pre,floatfix,twocolumn,showpacs]{revtex4}
\usepackage{amsmath}
\usepackage{amsfonts}
\usepackage{mathrsfs}
\usepackage{epsfig}

\begin{document}
\title{Modular networks with hierarchical organization: The dynamical
implications of complex structure}

\author{Raj Kumar Pan}
\author{Sitabhra Sinha}
\affiliation{%
The Institute of Mathematical Sciences, C.I.T. Campus, Taramani,
Chennai - 600 113 India
}%
\date{\today}
\begin{abstract}
Several networks occurring in real life have modular structures
that are arranged in an hierarchical fashion. In this paper, we have
proposed a model for such networks, using a stochastic generation
method. Using this model we show that, the scaling relation between the
clustering and degree of the nodes is not a necessary property of
hierarchical modular networks, as had previously been suggested on the basis
of a deterministically constructed model. We also look at dynamics on such
networks, in particular, the stability of equilibria of network
dynamics and of synchronized activity in the network. For both of these, we
find that, increasing modularity or the number of hierarchical levels tends
to increase the probability of instability. As both hierarchy and
modularity are seen in natural systems, which necessarily have to be robust
against environmental fluctuations, we conclude that additional constraints
are necessary for the emergence of hierarchical structure, similar to the
occurrence of modularity through multi-constraint optimization as shown by
us previously.
\end{abstract}
\maketitle

\section{Introduction}
\begin{figure}
\begin{center}
  \includegraphics[width=0.95\linewidth]{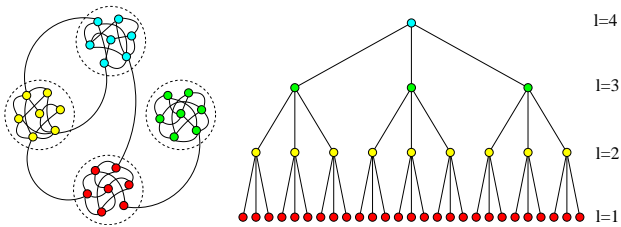}
\end{center}
\caption{
Schematic diagrams of (left) a modular network, with modules demarcated
by broken circles, and (right) a hierarchical network with 4 levels, each
indicated by a level number $l$.}
\label{fig:hie_mod}
\end{figure}

Structural patterns in complex networks occurring in biological, social and
technological contexts, have been a focus of study by physicists for a
decade, since the groundbreaking discovery of small-world property
\cite{Watts98} and scale-free degree distribution \cite{Barabasi99} for
many networks. One of the common features seen in many networks is the
occurrence of {\it modules}, namely, subnetworks whose members are highly
inter-connected but have few links to nodes outside the module. Many
networks have also been seen to have {\it hierarchical} organization, i.e.,
they are composed of successive interconnected layers or inter-nested
communities. In the literature, often the terms hierarchy and modularity
have been used almost inter-changeably, although, as shown in
Fig.~\ref{fig:hie_mod}, they represent distinct properties of the network.
However, it is interesting to note that these two properties have been
found to coexist in many networks occurring in real
life~\cite{Ravasz02,Sole04,Holme03,Krause03}, including the
Internet~\cite{Satorras01,Eriksen03} and the network of cortical areas in
the cat brain~\cite{Zhou06}.

Most of the complex systems seen in real life also have associated dynamics
\cite{Strogatz01}, and the structural properties of such networks have been
sought to be linked with their dynamical
behavior~\cite{Boccalett06,Barahona02}. In this respect, one of the
questions of obvious significance is whether there is a relation between
the {\em stability} of the dynamics against small perturbations in the
dynamical variables and the specific arrangement of the network's
connections. If the perturbation decays quickly, so that it is unable to
spread to the rest of the network, the network is said to be {\em stable}.
Such a property is necessary if networks are to survive the noisy
environment that characterizes the real world. It has sometimes been argued
that, networks with larger number of nodes, links and stronger 
inter-connections are
more stable. Such assertions are partly based on empirical observations,
e.g., in ecology, where it has been found that more diverse and strongly
connected ecosystems are more robust than their smaller, weakly connected
counterparts~\cite{Elton58}. On the other hand, theoretical work on the
stability of model networks have suggested the opposite conclusion. In
particular, according to the May-Wigner theorem~\cite{May73} for random
networks, increasing the complexity (as measured by the number of nodes,
density of connections and dispersion of interaction strengths) always
leads to decreased stability. However, this result is based on the study of
networks whose connection topology shows none of the structures that are
seen in real life networks, in particular, modularity and hierarchy.
Therefore, it is of interest to see whether introducing hierarchical
organization and modular structures can result in refutation of the
May-Wigner theorem. Early work on the stability of simple, structured model
networks~\cite{McMurtrie75} seemed to indicate that such structures indeed
promote stability, and this was also seen under certain conditions for
hierarchically organized networks~\cite{Hogg89}. However, a later study of
hierarchical, as well as, modular networks, concluded that these are less
stable than corresponding random networks~\cite{Hastings92}. We revisit
this problem in the present paper, by proposing a network model that
exhibits both modular structure and hierarchical organization. In addition
to looking at the stability of equilibria of the network
dynamics, we also consider the stability of synchronization over the
network. Although these two stability phenomena are superficially similar,
they involve looking at different properties of the network. The issue of
network synchronization, in particular, has assumed importance in recent
years, owing to its connection with, e.g., brain dynamics~\cite{Zhou06}.

An alternative model for hierarchical modular networks has been earlier
proposed by Ravasz and Barabasi (RB)~\cite{Ravasz03}. This model generates
a set of inter-nested modules in a hierarchical fashion using a
deterministic procedure that has both high clustering (because of the
modular nature of the network at the most fundamental level) and a
scale-free degree-distribution. These two properties do not always co-occur
in other network models that have been proposed in the literature. In
particular, the Barabasi-Albert (BA) network model~\cite{Barabasi99} allows
generation of a network with scale-free degree distribution through the
preferential attachment mechanism, but the average clustering coefficient of
its nodes decays with system size $N$.  Further, in the RB model, a scaling
relation is observed between the clustering coefficient of a node $C$ and
its number of connections (i.e., degree) $k$:
\begin{equation}
	C (k) \sim k^{-1}.
\end{equation}
Similar relations were also observed in several real networks, such as the
web of semantic connections between two English words which are
synonyms~\cite{Ravasz03}. This occurrence of the scaling relation between
clustering and degree of the nodes in a network has often been taken as a
signature for the existence of hierarchical modular structure in that
network. Recently, this scaling relation was shown to be actually an
outcome of degree-correlation bias in the usual definition of clustering
coefficient~\cite{Soffer05}.

However, it can be easily seen that this scaling relation is not a
necessary indicator for the existence of either modularity or hierarchy.
For example, consider a modular network consisting of $N$ nodes and $m$
modules of equal size. Let each node have degree $k$, with the links
initially occurring exclusively between nodes belonging to the 
same module (i.e., the
modules are isolated from each other). To make the network connected we
rewire a small fraction of the links keeping the degree of each node fixed.
Plotting clustering as a function of degree for this network will only show
vertical spread of points at a single node degree value. Let us consider
another example, this time a hierarchical structure, viz., the Cayley tree
with $b$ branches at each vertex. Again, it is easy to see that the
clustering versus degree curve will not show the characteristic scaling
seen for the RB model. In fact, in the next section, we show that even for
networks where both hierarchy and modularity are present, it is not necessary
that this scaling relation between clustering and node degree will hold.

The paper is organized as follows. In the next section we introduced a
simple model of a modular network with hierarchical organization. In
section~\ref{sec:dynamics} we introduce the formalism to analyze the
stability of dynamical equilibria and synchronized states of a network.
The proposed model allows a detailed study of the relation between
dynamical stability and hierarchical modular organization of the
network. We observe that both of these structural properties actually
increase the instability compared to an equivalent random network. This
may appear counter-intuitive as both modularity and hierarchy are observed
in networks occurring in nature, which necessarily have to be robust to
survive environmental fluctuations. However, the emergence of modular
structures can be understood as a response to multiple (and often
conflicting) constraints imposed on such networks~\cite{Pan07}. We
conclude with a discussion about how these observations can possibly be
extended to explain the emergence of hierarchical organization.

\section{Model of Hierarchical Modular Network}
\label{sec:model}
\begin{figure}
\begin{center}
  \includegraphics[width=0.95\linewidth]{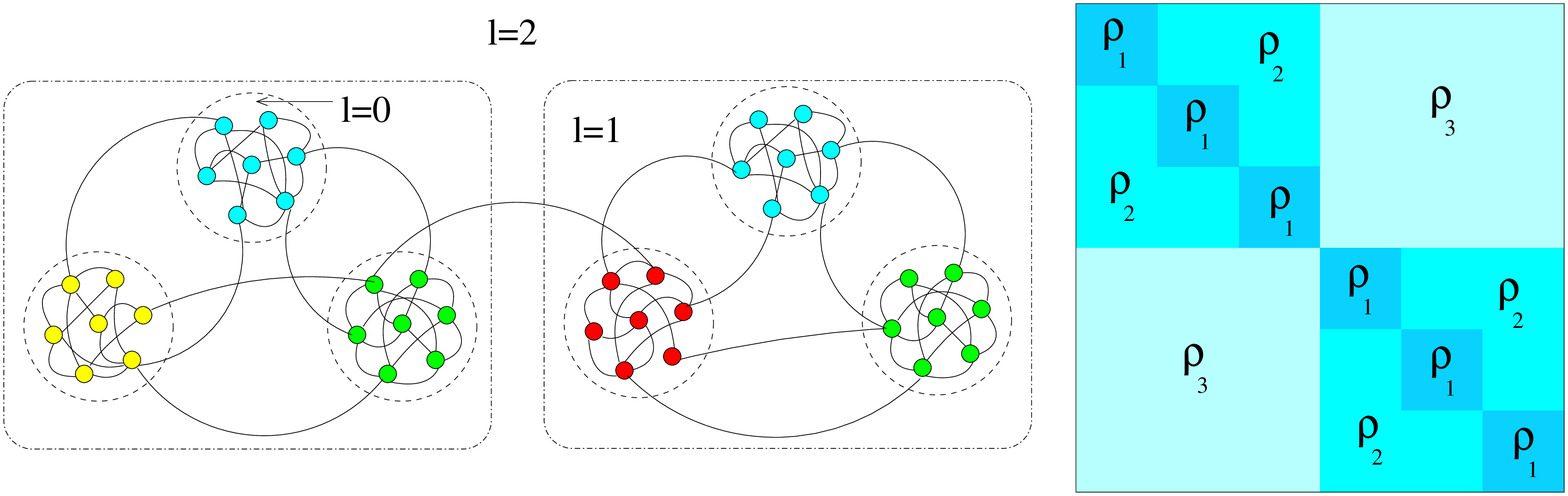}
  \end{center}
\caption{Schematic diagram of the hierarchical modular network model (left)
with the modules occurring at the various hierarchical levels ($l$)
indicated by broken lines, and the corresponding adjacency matrix (right)
where $\rho_1$ indicates the density of connections within and, $\rho_{l+1}$,
between the different modules at each level $l$.}
\label{fig:model}
\end{figure}
Here we propose a general model for networks having modular as well as
hierarchical structure. Let us begin with a modular network consisting of
$m$ modules, each containing $n$ nodes. The connectivity (i.e., the
probability of a link between any pair of nodes) within each module is
$\rho_{1}$, while the connectivity between modules is $\rho_{2}$ ($\leq
\rho_{1}$). We now introduce hierarchy by adding another set of $m$ modules
(each having $n$ nodes) with the same $\rho_{1}$ and $\rho_{2}$. The nodes
belonging to these two different sets of modules are now connected, but
with a probability $\rho_{3}$ ($\leq \rho_{2}$). The resulting network has
$2 n m$ nodes and $l=2$ hierarchical levels (Fig.~\ref{fig:model}). To
increase the number of hierarchical levels to $l=3$, we add a similar
network with $2n m$ nodes to the existing network and, as above, add links
between these two networks with a probability $\rho_{4}$ ($\leq \rho_{3}$).
Thus, to get a network with $l=h$ hierarchical levels, the above procedure
is repeated $h-1$ times. The final network contains $M=2^{h-1} m$
number of modules. Note that, all connections between nodes are made
randomly. To reduce the number of model parameters, we assume that the
connectivities $\rho_1,\ldots,\rho_{h+1}$ are related as: 
\begin{equation}
\frac{\rho_2}{\rho_1}=\frac{\rho_3}{\rho_2}=\cdots=\frac{\rho_{h+1}}{\rho_{h}}=r,
\end{equation}
where, $0\leq r \leq 1$, the ratio of inter-modular connections between
two successive hierarchical levels, is a control parameter. 
By varying $r$, one can
switch between isolated modular ($r = 0$) and homogeneous random ($r = 1$)
networks, with intermediate values of $r$ giving hierarchical modular
networks. We compare between networks having different number of
hierarchical levels $h$, keeping the total number of modules $M$ and
average degree $\langle k \rangle$ fixed.

To consider the effect of hierarchy in isolation, while keeping modularity
fixed (e.g., as measured by the Newman modularity measure $Q$ \cite{Newman04a}),
we use a variant of the above model, where, $\rho_{1}$~=~constant, while other connectivities are still related by
\begin{equation}
	\frac{\rho_3}{\rho_2}=\cdots=\frac{\rho_{h+1}}{\rho_{h}}=r.
\end{equation}
This implies that the average number of intra-modular ($\langle
k_{\text{intra}} \rangle$) 
and inter-modular ($\langle k_{\text{inter}} \rangle$) connections per
node are also constant 
\footnote{Note that, $\langle k_{\text{intra}} \rangle = \rho_{1}
\left(\frac{N}{M}-1 \right)$, and \\
$\langle k_{\text{inter}} \rangle = N\rho_{2}\left[ \frac{(m-1)}{M} +
r\left(\frac{1}{2}\right)^{h-1}
+ \cdots+r^{h-1}\left(\frac{1}{2}\right)\right]$.}. 

\begin{figure}
\begin{center}
  \includegraphics[width=0.8\linewidth]{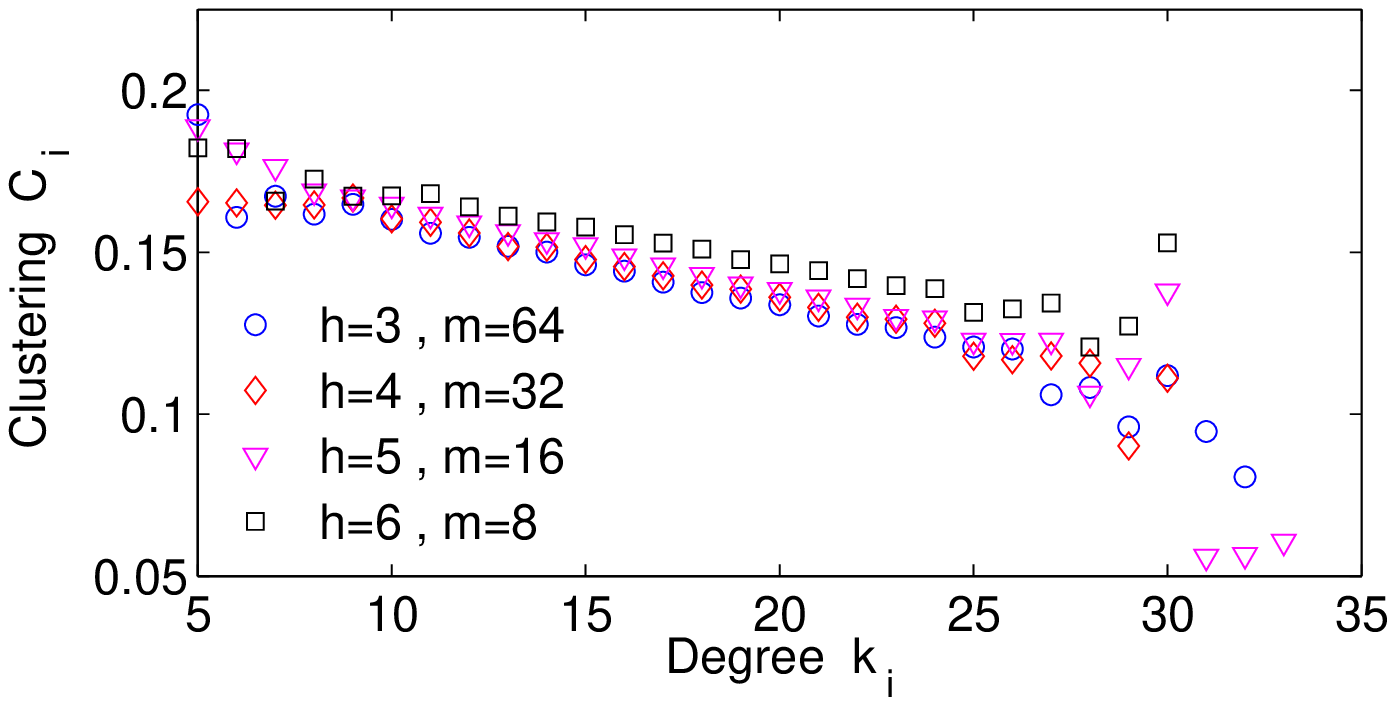}
	\includegraphics[width=0.8\linewidth]{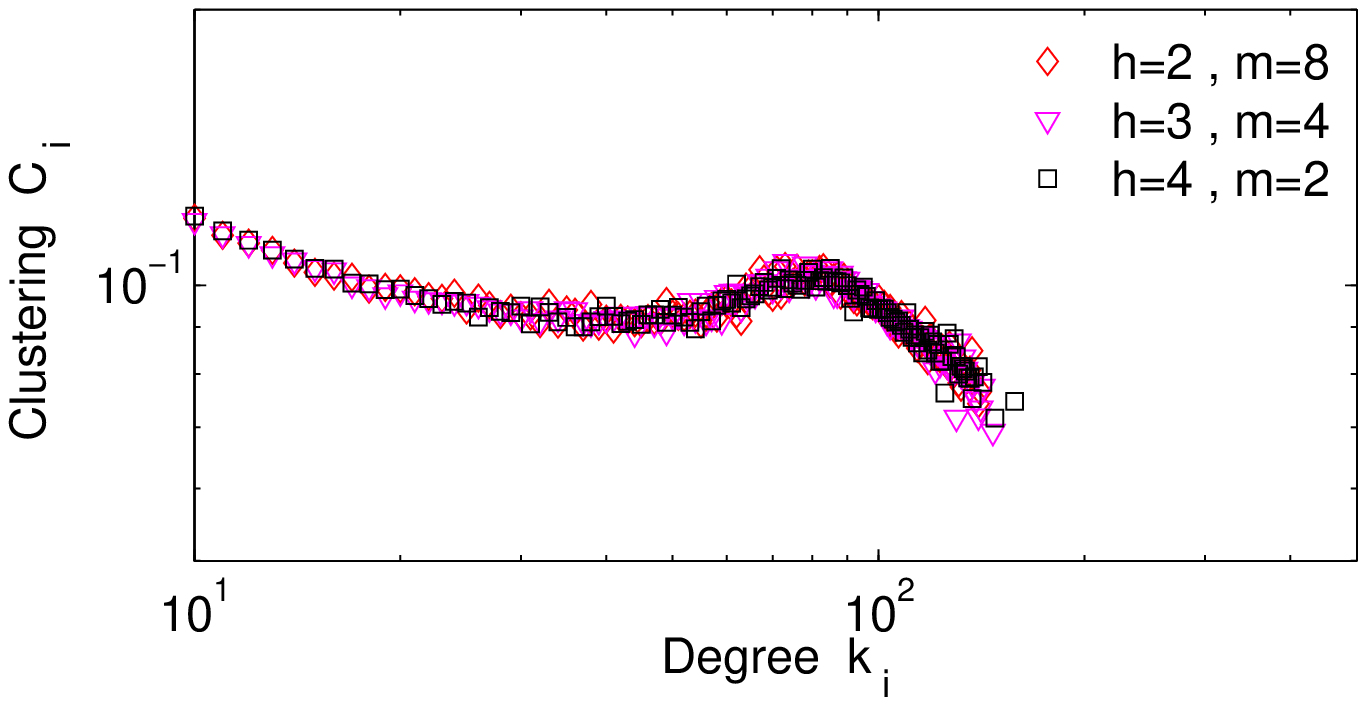}
\end{center}
\caption{Clustering coefficient $C_i$ of the $i$-th node as a function of
its degree $k_i$ for the hierarchical modular network model proposed here,
where each module at $l=1$ is (left) a random ER network and (right) a
scale-free BA network. The different symbols indicate networks with
differing total number of hierarchical levels, $h$. For both types of
networks, the total number of nodes is $N=8192$ with average intra-modular
degree, $\langle k_{\text{intra}} \rangle = 10$, inter-modular degree,
$\langle k_{\text{inter}}\rangle=5$, and the ratio of inter-modular
connections between two successive hierarchical levels, $r=0.1$. Note that,
in neither case is a scaling relation observed between $C_i$ and $k_i$,
although the modules are arranged in a hierarchical manner by
construction.}
 \label{fig:scaling}
\end{figure}
The stochastic construction procedure of this network, along with the
ability to vary modularity (by changing $r$) independently of the number of
hierarchical levels ($h$), makes it an extremely general model. In
addition, as it is hierarchical by construction, we can show that the
criterion suggested in Ref.~\cite{Ravasz03}, namely, the scaling relation
between clustering and degree, is not a necessary condition for the
existence of hierarchical modularity. As shown in
Fig.~\ref{fig:scaling}~(left), when the modules are random networks, the
scaling relation is clearly absent for our model network. To counter the
possible argument that this failure of the relation is due to the
non-scale-free degree distribution, we have also considered the case where
each of the modules is a BA network. Although the inter-modular
connections are made randomly, the network degree distribution is still
scale-free. Even for this case, a clear scaling relation between clustering
and degree is absent (Fig~\ref{fig:scaling}, right).

\section{Dynamics on Hierarchical Networks}
\label{sec:dynamics}
\subsection{Linear Stability of Equilibria}
To look at the effect of hierarchy on network dynamics, we consider the
linear stability of an arbitrarily chosen equilibrium state for a 
set of coupled
differential equations defining the time-evolution of the system. For a
network of $N$ nodes, a dynamical variable $x_i$ is associated with each
node $i$. The state of the system, ${\bf x}$, can be characterized by
$\dot{\bf x} = f({\bf x})$, where $f$ is a general nonlinear function. To
investigate the stability around an arbitrary fixed point ${\bf {x}^*}$
(i.e., $f({\bf x})|_{\bf x^*}=0$), we check whether a small perturbation
$\delta{\bf x}$ about ${\bf {x}^*}$ grows or decays with time. This
perturbation evolves as
\begin{equation}
	\dot{\delta{\bf x}}={\bf J}{\bf x},
\end{equation}
where, ${\bf J}$ is the Jacobian matrix representing the interactions among
the nodes: $J_{ij}=\partial f_i/\partial x_j |_{\bf{x}^*}$. As we are
interested in the instability induced through the connections of the
network, rather than the intrinsic instability of individual unconnected
nodes, we can (without much loss of generality) set the diagonal element
$J_{ii}=-1$. This implies that, in the absence of any connections, the
nodes are self-regulating, i.e., the fixed point $\bf{x}^*$ is stable. The
behavior of the perturbation is determined by the largest real part,
$\lambda_{\text{max}}$, of the eigenvalues of ${\bf J}$. If
$\lambda_{\text{max}} > 0$, an initially small perturbation will grow
exponentially with time, and the system will be rapidly dislodged from the
equilibrium state ${\bf {x}^*}$.

The relation between the dynamical properties and the static structure of
the network is provided by its adjacency matrix ${\bf A}$ (with $A_{ij}=1$,
if nodes $i$ and $j$ are connected, and $0$ otherwise). There is a direct
correspondence between the nature of the matrices ${\bf J}$ (specifying the
dynamical behavior of perturbation) and ${\bf A}$ (which determines the
structure of the underlying directed network), because $A_{ij}= 0$ implies
$J_{ij}= 0$. In our model, we have generated $J_{ij}$ by randomly choosing
the non-zero elements from a Gaussian distribution with zero mean and
variance $\sigma^2$. For Erdos-Renyi (ER) random networks, ${\bf J}$ is an
unstructured random matrix and the largest real part of its eigenvalues,
$\lambda_{\text{max}} \sim \sqrt{N \rho \sigma^2}-1$, where $\rho$ is the
connectivity of the network, and $\sigma$ measures the dispersion of
interaction strengths~\cite{May73}. When any of the parameters, $N$,
$\rho$, or $\sigma$, is increased, there is a transition from stability to
instability. The critical value at which the transition to instability
occurs is $\sigma_c \sim 1/ \sqrt{N \rho}$. This result, implying that
complexity promotes instability, has been shown to be remarkably robust
with respect to various
generalizations~\cite{Jirsa04,Sinha05,Sinha05a,Sinha06}. 

\begin{figure}
\begin{center}
  \includegraphics[width=0.8\linewidth]{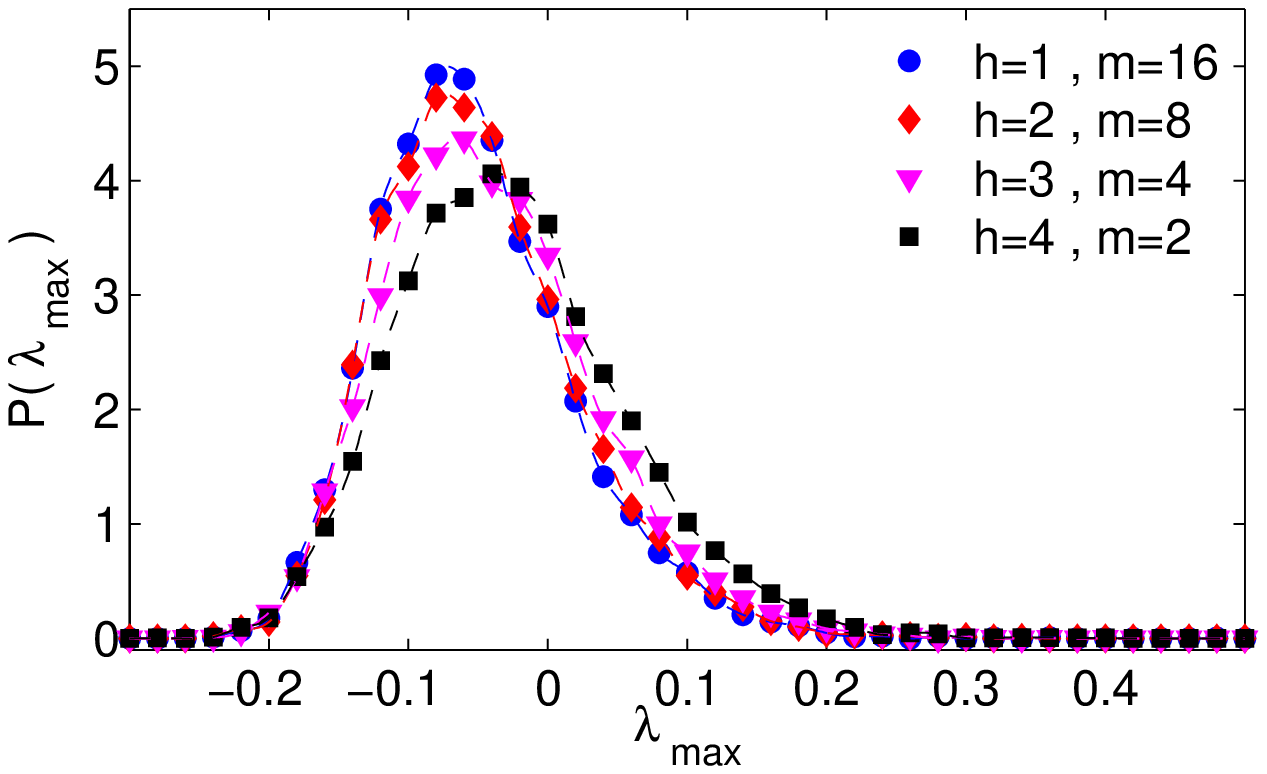}
  \includegraphics[width=0.8\linewidth]{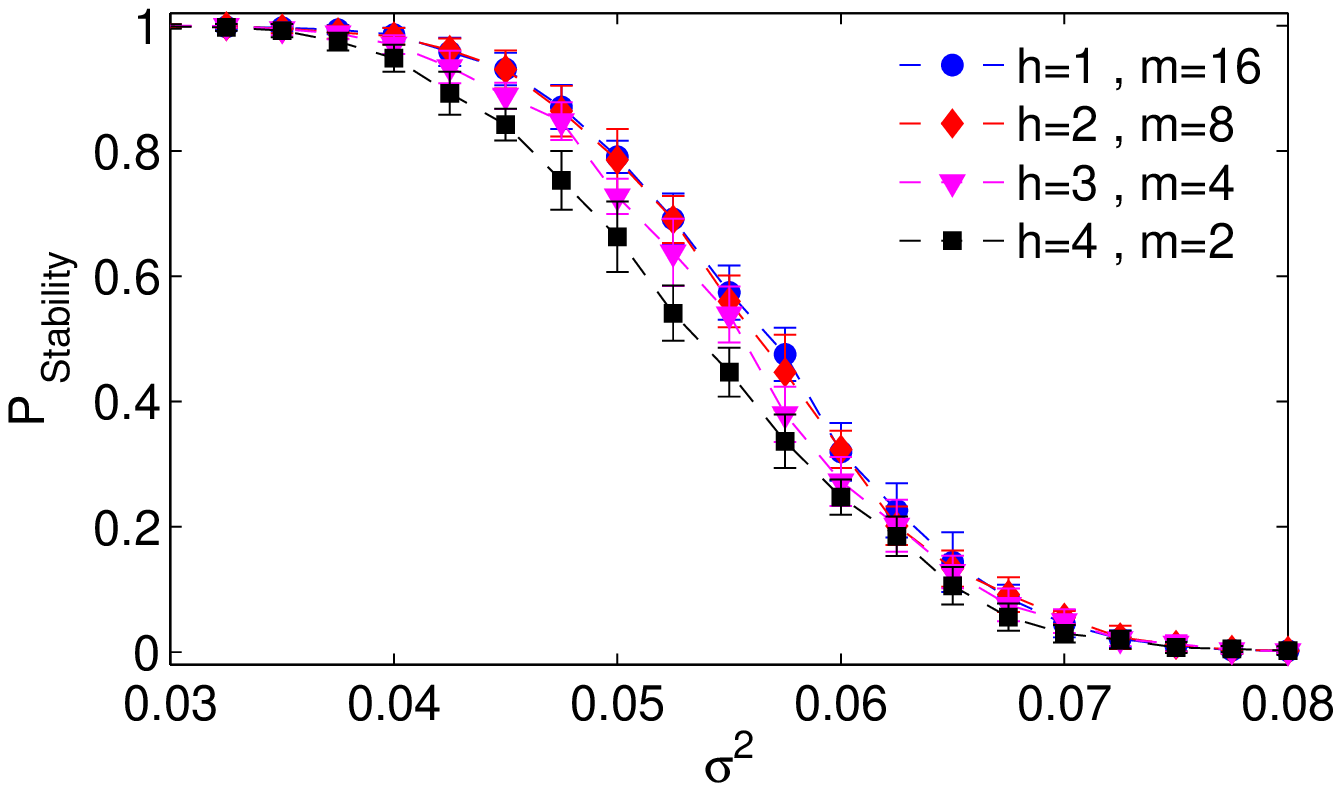}
\end{center}
\caption{(Left) Probability distribution for the largest real part of the
eigenvalues of the Jacobian $J$, as a function of total number of
hierarchical levels, $h$ (the interaction strength parameter, $\sigma^2 =
0.05$). (Right) Probability of stability for a hierarchical modular network
as a function of $\sigma^2$, with different symbols corresponding to
differing total number of hierarchical levels $h$. Link weights are chosen
from a $normal(0,\sigma^2)$ distribution. For all cases, the network
consists of $N = 256$ nodes with average intra-modular degree, $\langle
k_{\text{intra}} \rangle = 10$, inter-modular degree, $\langle
k_{\text{inter}}\rangle=5$, and the ratio of inter-modular connections
between two successive hierarchical levels, $r=0.1$. At all hierarchical
levels $l > 1$, the network is split into two sub-networks. At $l = 1$,
each subnetwork is split into $m$ modules ($l = 0$). Thus, $N = 256$
nodes are divided equally among $2^{h-1} m = 16$ modules, with the
four curves corresponding to ($\Box$) $h=4$, $m=2$, ($\bigtriangledown$)
$h=3$, $m=4$, ($\diamond$) $h=2$, $m=8$, and ($\circ$) $h=1$, $m=16$. Note
that, increasing $h$ causes the transition to instability to occur at a
smaller value of $\sigma^2$, implying that increasing hierarchy increases
instability.}
\label{fig:rand_hier}
\end{figure}
Here, using the above formalism, we examine the effect of hierarchy on the
stability of equilibria when one of the network
parameters (namely, $\sigma$) is varied. We study the critical value at
which the transition to instability occurs, $\sigma_c$, as a function of the
total number of hierarchical levels, $h$, keeping the total number of
modules $M$ fixed. We find that, with increasing $h$, the distribution of
$\lambda_{max}$ shifts towards more positive values
(Fig.~\ref{fig:rand_hier}, left). As the system becomes unstable when
$\lambda_{max} > 0$, it follows that the probability of stability 
for the network decreases
with increasing number of hierarchical levels (Fig.~\ref{fig:rand_hier},
right).

\subsection{Synchronization}
\begin{figure}
\begin{center}
  \includegraphics[width=0.8\linewidth]{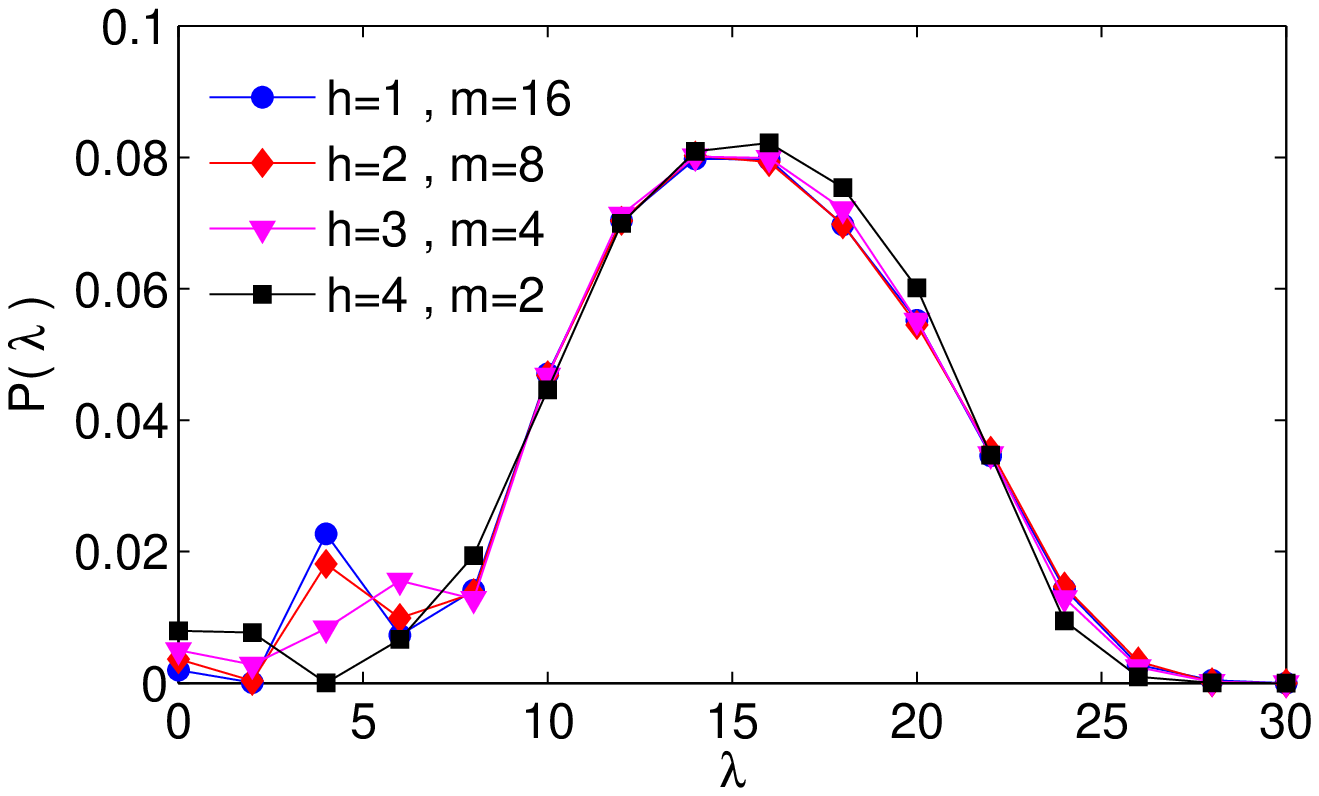}
  \includegraphics[width=0.8\linewidth]{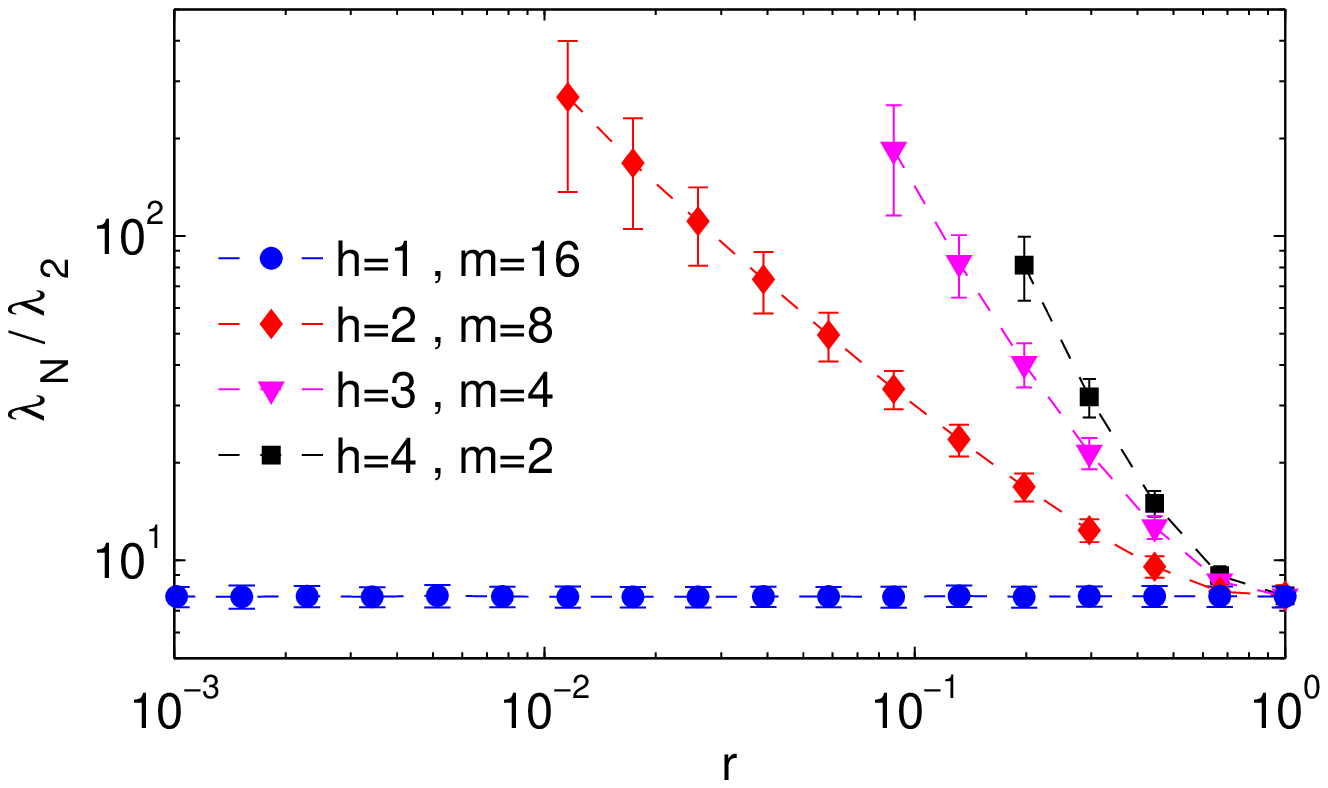}
  \end{center}
\caption{(Left) Probability distribution of eigenvalues of the Laplacian
$L$, as a function of the total number of hierarchical levels, $h$ ($r = 0.1$).
(Right) The ratio of the largest eigenvalue ($\lambda_N$) to the second
smallest eigenvalue ($\lambda_2$) as a function of $r$, the ratio of 
inter-modular
connections between two successive hierarchical levels, with different
symbols corresponding to differing total number of hierarchical levels $h$.
For all cases, the network consists of $N = 256$ nodes with average
intra-modular degree, $\langle k_{\text {intra}} \rangle = 10$ and
inter-modular degree, $\langle k_{\text {inter}}\rangle=5$. At all
hierarchical levels $l > 1$, the network is split into two sub-networks. At
$l = 1$, each subnetwork is split into $m$ modules ($l = 0$). Thus, $N
= 256$ nodes are divided equally among $2^{h-1} m = 16$ modules,
with the four curves corresponding to ($\Box$) $h=4$, $m=2$,
($\bigtriangledown$) $h=3$, $m=4$, ($\diamond$) $h=2$, $m=8$, and ($\circ$)
$h=1$, $m=16$. Note that, increasing the number of hierarchical levels leads
to divergence of the eigenratio, implying that synchronization becomes
harder to achieve.}
\label{fig:synch}
\end{figure}
It is of interest to look not only at the stability of equilibria for
network dynamics, but also at the stability of synchronized activity in
networks. Let us consider a network of $N$ identical oscillators. The
time-evolution of this coupled dynamical system is described by:
\begin{equation}
\dot{x}_{i}=\mathrm{F}(x_{i})+\epsilon\sum_{j=1}^{n}L_{ij}\mathrm{H}(x_{j}).
\label{eq:variational}
\end{equation}
Here, $x_{i}$ is a variable associated with node $i$; $F$ and $H$ are
evolution and output functions, respectively; $\epsilon$ is the strength of
coupling; and ${\bf L}$ is the Laplacian matrix, defined as:
$L_{ii}=k_{i}$, the degree of node $i$, $L_{ij}=-1$ if nodes $i$ and $j$
are connected, $0$ otherwise. It has been shown that the linear stability
of the synchronized state $x_{s}$ (=$x_1=\ldots=x_N$) can be determined by
diagonalizing the variational equation (Eq.~\ref{eq:variational}) into $N$
blocks of the form, $\dot{y_{i}}=[DF(s)+\epsilon\lambda_{i}DH(s)]y_{i}$,
where $y_i$ represent different modes of perturbation from the synchronized
state. This is also referred to as the \textit{master stability
equation}~\cite{Barahona02}. These equations have the same form but
different effective couplings $\alpha_i = \epsilon \lambda_{i}$.  The
synchronized state is stable, i.e., the maximum Lyapunov exponent is in
general negative, only within a bounded interval $[\alpha_{A},\alpha_{B}]$
\cite{Pecora98}. Let the eigenvalues of the Laplacian matrix be arranged as
$0=\lambda_{1}<\lambda_{2}\leq \dots \leq \lambda_{n}$. Then, requiring all
effective couplings to lie within the interval $\alpha_{A} <
\epsilon\lambda_{2} \leq \dots \leq \epsilon\lambda_{N}<\alpha_{B}$,
implies that a synchronized state is linearly stable, if and only if,
$\lambda_{N}/\lambda_{2}<\alpha_{B}/\alpha_{A}$. Thus, a network having a
smaller eigenratio $\lambda_{N}/\lambda_{2}$, is more likely to show stable
synchronized activity.

Here, we obtain the eigenvalues of the Laplacian ${\bf L}$ for a
hierarchical modular network (Fig.~\ref{fig:synch}, left) and observe the
eigenratio $\lambda_{N}/\lambda_{2}$ as a function of ratio of the
inter-modular connections between two successive hierarchical levels, $r$,
and the total number of hierarchical levels, $h$. First, keeping the number
of hierarchical levels fixed, we vary the parameter $r$. We find that with
decreasing $r$, i.e., as the number of connections between two successive
hierarchical
levels decrease, the instability of the synchronized state increases.
Next, keeping the total number of modules fixed we increase the number of
hierarchical levels ($h$) in the network. Fig.~\ref{fig:synch} (right)
shows that as the number of hierarchal levels of the network is increased,
$\lambda_{2}$ decreases, resulting in an increasing eigenratio. Thus,
arranging the modules of a network in a hierarchical fashion also makes a
network difficult to synchronize.

\section{Discussion and Conclusion}
\begin{figure}
	\begin{center}
		\includegraphics[width=0.95\linewidth]{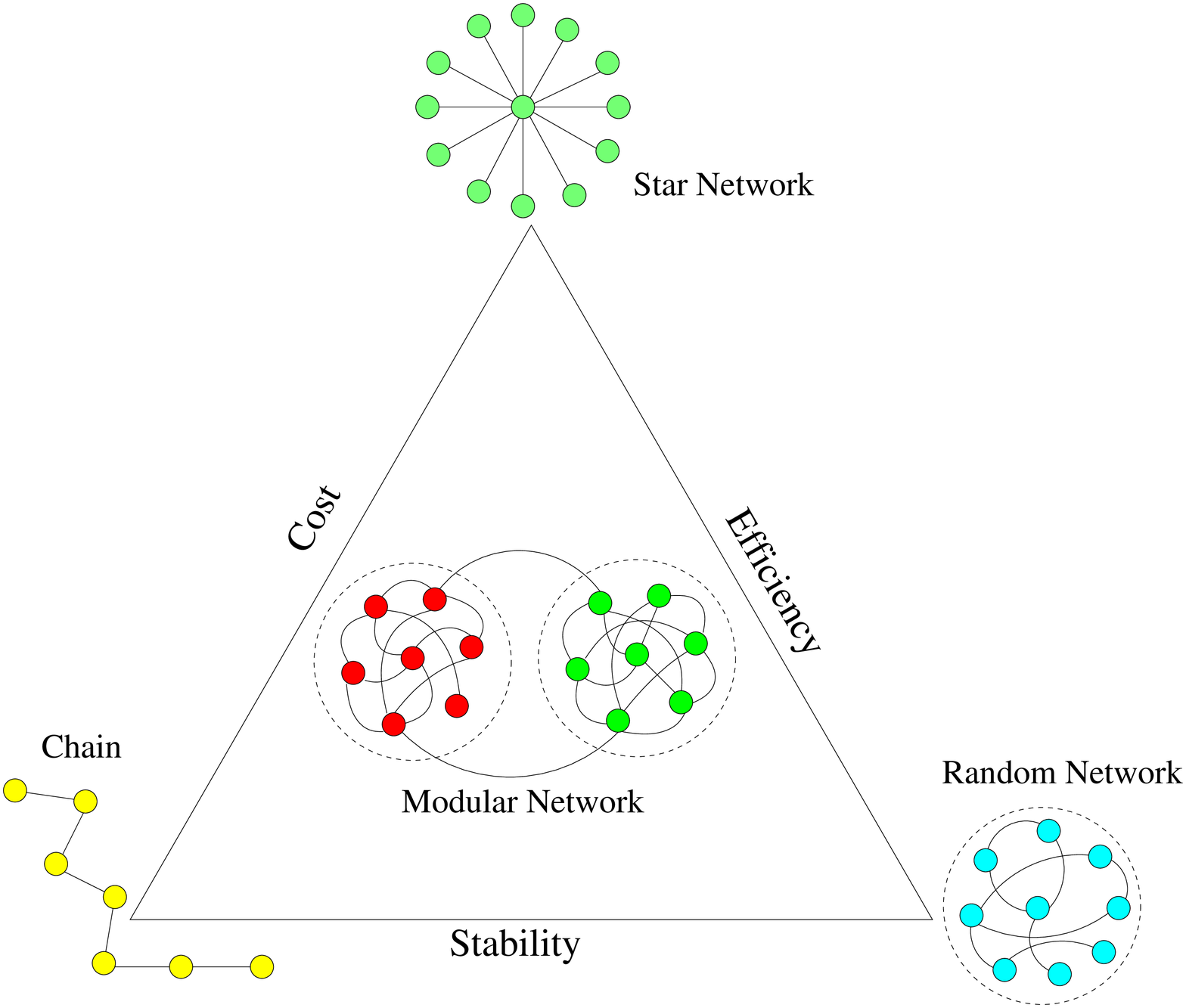}
	\end{center}
\caption{Schematic diagram indicating the different types of optimal
networks obtained by satisfying different constraints. Each vertex
represents networks obtained by satisfying a pair of constraints. Note
that, modular networks emerge by optimizing all three constraints, viz.,
cost, efficiency and stability, indicated by the three arms of the
triangle.}
\label{fig:schematic_mod}
\end{figure}
In previously published work~\cite{Pan07}, we have shown that increased
modularity in random networks leads to higher probability of instability
for the equilibria of the network dynamics. Thus, the work presented here
is an extension and generalization of the above result, demonstrating that
increased number of hierarchical levels also tend to destabilize these
equilibria, and moreover, the same phenomena is observed for the stability
of synchronized activity in a network with respect to increased modularity
and hierarchy. This raises the question of how can systems with
hierarchical modular structures be seen in nature at all, where they have
to be robust enough to survive constant environmental fluctuations. An
answer to this can be fashioned along the lines of our recent work showing
that additional natural constraints operating on networks in real life,
such as the minimization of (a)~resource cost for maintaining each link and
(b)~the time required for communicating between nodes, in addition to
linear stability of equilibria, will make modular networks the optimal
configuration (Fig.~\ref{fig:schematic_mod})~\cite{Pan07}. 
We find such stable, modular networks to
possess multiple hubs and a heterogeneous degree distribution. Many types
of networks, including scale-free networks~\cite{Barabasi99}, can be seen
as special cases of this general criterion. Therefore, we can understand
the large-scale occurrence of such networks in nature as a response to
co-existing structural and dynamical constraints. 

One can ask, what will be the effect of introducing constraints other than
the ones mentioned here. For example, replacing the criterion for linear
stability by one demanding robustness with respect to removal of links
(selected by using a combination of random and targeted attack strategies)
does not qualitatively change our results. It turns out that this criterion
is satisfied by networks with bimodal degree distribution, a property that
our optimal modular networks possess. However, while this can explain the
ubiquity of modularity, it does not answer the question of why hierarchical
organization is so common in nature. The fact that tree-like networks with
extensive ramifications occur so often in the context of resource transport
(e.g., the circulatory system in plants and animals) suggest that
additional constraints related to flow maximization may be at work in this
case. Another possible candidate for such a constraint may be the need to
minimize wiring cost, i.e., the total link length~\cite{Mathias01}. This is
applicable when the network is embedded on a geographic (as opposed to
topological) space, so that the wiring cost can been defined as the sum of
the Euclidean distances between all connected pairs of nodes. As many of
the networks showing hierarchical organization (such as the internet and
the network of cortical areas in the brain) are indeed defined in metric
space, this is a possibility that needs to be analysed in detail.

\end{document}